\documentclass[aps,prl,amsmath,amssymb,amstext,citeautoscript,punctuation,twocolumn,nofootinbib,superscriptaddress]{revtex4-2}
\usepackage[dvips]{graphicx}
\usepackage{amsmath,amssymb,amsthm,mathrsfs,amsfonts,dsfont,subfigure,epsfig,braket,bm,enumerate,color,graphicx,dcolumn}
\usepackage[normalem]{ulem}
\usepackage[dvipsnames]{xcolor}
\usepackage{xcolor,charter, gensymb, float}
\usepackage [english]{babel}
\usepackage{xspace}
\DeclareMathOperator{\Tr}{Tr}
\usepackage{siunitx}
\usepackage{bm}
\graphicspath{{FiguresMain/}}

\newcommand{\bmro}{Ba$_2$MgReO$_6$\xspace}
\newcommand{\kQ}{^{\bm k}_\theta}

\newcommand{\vk}{{\bm{k}}}
\DeclareSIUnit\angstrom{\text {Å}}

\begin{document}
\title{Spectroscopic signatures and origin of hidden order in Ba$_2$MgReO$_6$}
\author{Jian-Rui Soh}%
\thanks{These authors contributed equally to this work.}
\affiliation{Institute  of  Physics,  \'Ecole  Polytechnique  F\'ed\'erale  de  Lausanne  (EPFL),  CH-1015  Lausanne,  Switzerland}
\author{Maximilian E. Merkel}
\thanks{These authors contributed equally to this work.}
\affiliation{Materials Theory, ETH Z\"u{}rich, Wolfgang-Pauli-Strasse 27, 8093 Z\"u{}rich, Switzerland}
\author{Leonid V. Pourovskii}%
\affiliation{CPHT, CNRS, École Polytechnique, Institut Polytechnique de Paris, 91120 Palaiseau, France}%
\affiliation{Collège de France, Université PSL, 11 place Marcelin Berthelot, 75005 Paris, France}
\author{Ivica Živković}
\affiliation{Institute  of  Physics,  \'Ecole  Polytechnique  F\'ed\'erale  de  Lausanne  (EPFL),  CH-1015  Lausanne,  Switzerland}
\author{Oleg Malanyuk}
\affiliation{Institute  of  Physics,  \'Ecole  Polytechnique  F\'ed\'erale  de  Lausanne  (EPFL),  CH-1015  Lausanne,  Switzerland}
\author{Jana Pásztorová}%
\affiliation{Institute  of  Physics,  \'Ecole  Polytechnique  F\'ed\'erale  de  Lausanne  (EPFL),  CH-1015  Lausanne,  Switzerland}%
\author{Sonia Francoual}
\affiliation{Deutsches Elektronen-Synchrotron DESY, Notkestraße 85, 22607 Hamburg, Germany}%
\author{Daigorou Hirai}%
\affiliation{Department of Applied Physics, Nagoya University, Nagoya 464-8603, Japan}
\author{Andrea Urru}
\affiliation{Materials Theory, ETH Z\"u{}rich, Wolfgang-Pauli-Strasse 27, 8093 Z\"u{}rich, Switzerland}
\author{Davor Tolj}
\affiliation{Institute  of  Physics,  \'Ecole  Polytechnique  F\'ed\'erale  de  Lausanne  (EPFL),  CH-1015  Lausanne,  Switzerland}
\author{Dario Fiore Mosca}%
\affiliation{CPHT, CNRS, École Polytechnique, Institut Polytechnique de Paris, 91120 Palaiseau, France}%
\affiliation{Collège de France, Université PSL, 11 place Marcelin Berthelot, 75005 Paris, France}
\author{Oleg V. Yazyev}
\affiliation{Institute  of  Physics,  \'Ecole  Polytechnique  F\'ed\'erale  de  Lausanne  (EPFL),  CH-1015  Lausanne,  Switzerland}%
\author{Nicola A. Spaldin}
\affiliation{Materials Theory, ETH Z\"u{}rich, Wolfgang-Pauli-Strasse 27, 8093 Z\"u{}rich, Switzerland}
\author{Claude Ederer}
\affiliation{Materials Theory, ETH Z\"u{}rich, Wolfgang-Pauli-Strasse 27, 8093 Z\"u{}rich, Switzerland}
\author{Henrik M. Rønnow}
\affiliation{Institute  of  Physics,  \'Ecole  Polytechnique  F\'ed\'erale  de  Lausanne  (EPFL),  CH-1015  Lausanne,  Switzerland}%
\date{\today}
\maketitle
\textbf{Clarifying the underlying mechanisms that govern ordering transitions in condensed matter systems is crucial for comprehending emergent properties and phenomena. While transitions are often classified as electronically driven or lattice-driven, we present a departure from this conventional paradigm in the case of the double perovskite \bmro{}.  Leveraging resonant and non-resonant elastic x-ray scattering techniques, we unveil the simultaneous ordering of structural distortions and charge quadrupoles at a critical temperature of $T_\mathrm{q}$$\sim$$\SI{33}{K}$. Using a variety of complementary first-principles-based computational techniques, we demonstrate that, while electronic interactions drive the ordering at $T_\mathrm{q}$, it is ultimately the lattice distortions that dictate the specific ground state that emerges. Our findings highlight the crucial interplay between electronic and lattice degrees of freedom, providing a unified framework to understand and predict unconventional emergent phenomena in quantum materials.}

\section{Introduction}

Understanding symmetry-lowering phase transitions emerging from the ordering of electronic charge, orbital, magnetic, and/or structural degrees of freedom represents a fundamental objective of condensed matter physics. Recently, various materials containing 4$d$ or 5$d$ transition-metal cations have attracted considerable attention~\cite{Pasztorova2022, Hirai2019, Hirai2020, Erickson2007, Lu2017, 5d1_magnetic_order_muSR, xu_covalency_2016, PhysRevResearch.5.L012010, PhysRevB.104.174422, PhysRevB.100.041108, Barbosa, Marjerrison, YAMAMURA2006605, PhysRevB.100.045142, frontini_spin-orbit-lattice_2023}, owing to the comparable strengths of interactions which drive transitions. In particular, the size of the spin-orbit coupling, intersite electron hopping energies, and crystal-field effects in these materials are finely balanced, resulting in strongly entangled spin and orbital degrees of freedom that support exotic forms of order, such as higher-order magnetic or charge multipoles~\cite{Chen2010, 5d2_chen_gang, 5d1_review_cooperation, 5d1_Mean_field_orbital_order}.

\begin{figure}[b!]
\centering
\includegraphics[width=0.99\columnwidth]{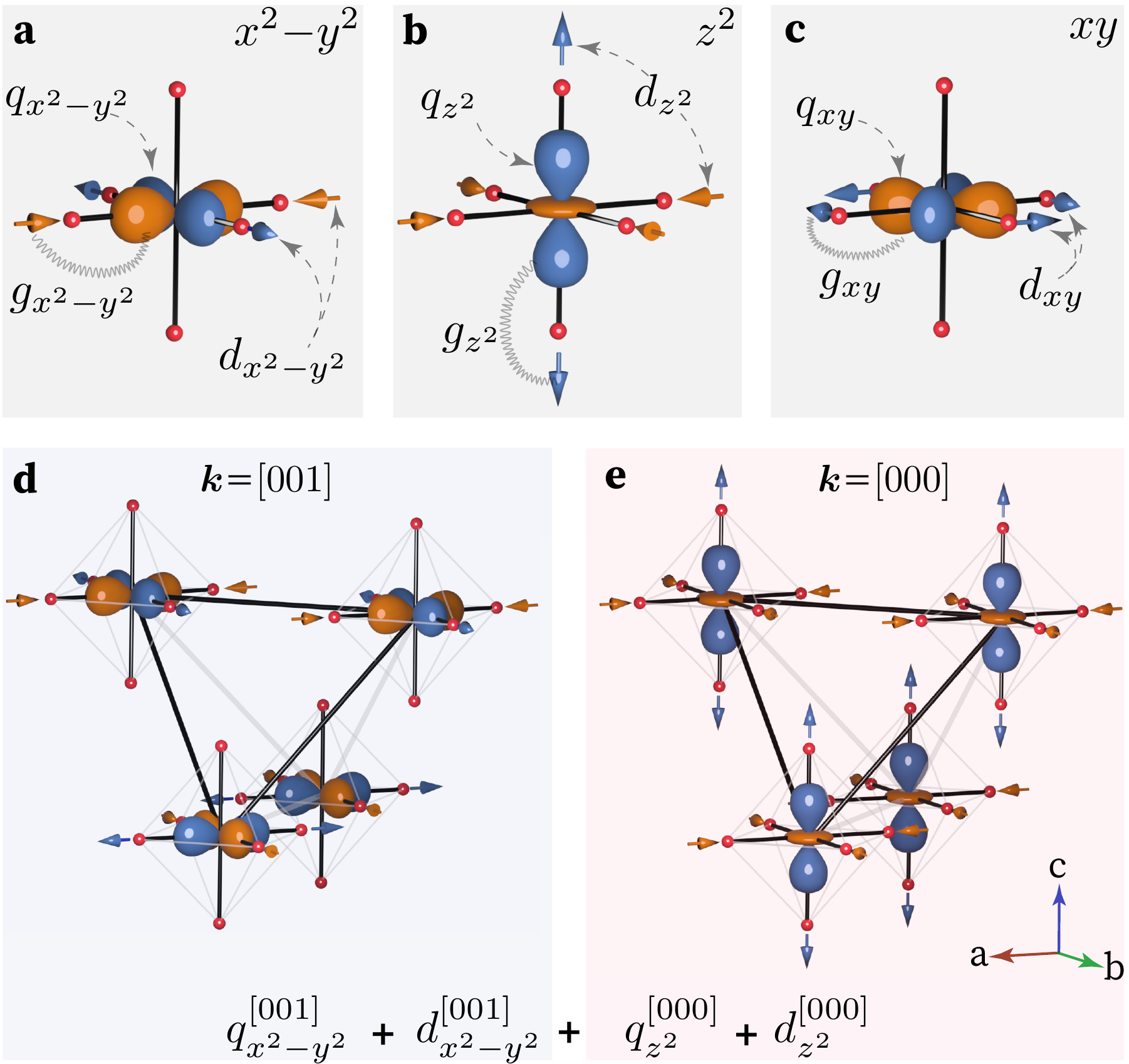}
\caption{\textbf{Lattice and electric-quadrupolar coupling and ordering}. \textbf{a}-\textbf{c} The electron-lattice coupling ($g_\theta$) between lattice deformations ($d_{\theta}$) and charge quadrupoles ($q_{\theta}$) with symmetry $\theta$=$x^2-y^2$, $z^2$ and $xy$, respectively. The blue and orange lobes of $q_{\theta}$ correspond to regions with excess and reduced electronic charge, respectively. \textbf{d}, \textbf{e} Below $T_\mathrm{q}$, BMRO develops a simultaneous long-range order of the local distortion $d\kQ$ and electric quadrupoles $q\kQ$. \textbf{d} The $\theta$$=$$x^2$$-$$y^2$ component exhibit an antiferroic order with a $\vk$$=$$[001]$ propagation vector. \textbf{e} The $\theta$$=$$z^2$ component display ferroic, $\vk$$=$$[000]$ arrangement.}
\label{fig:FIG1}
\end{figure}

However, the direct detection of such multipolar order is challenging, and thus it is sometimes described as a ``hidden'' order. For example, the ordering of charge quadrupoles in several $5d^1$ and $5d^2$ double perovskite (DP) systems predicted by various theoretical studies~\cite{5d1_Mean_field_orbital_order, Chen2010, LoveseyBMRO, iwahara_vibronic_2023} has eluded experimental verification thus far. Moreover, the emergence of multipolar order can be accompanied by structural distortions, which are generally easier to detect experimentally. However, in such cases, it is often unclear whether the initial driving force of the transition is purely electronic, and the structural distortion is merely a by-product of the electronic symmetry-lowering, or whether the structural distortion enables the transition in the first place. Furthermore, the boundaries between these categories are not always clear-cut, as numerous examples exhibit intricate phenomena stemming from the interplay between electronic and lattice interactions~\cite{peil_mechanism_2019, georgescu_quantifying_2022, pavarini_mechanism_2008, pavarini_origin_2010, zhang_lavo3_2022}.

\begin{figure*}[ht!]
\centering
\includegraphics[width=1.99\columnwidth]{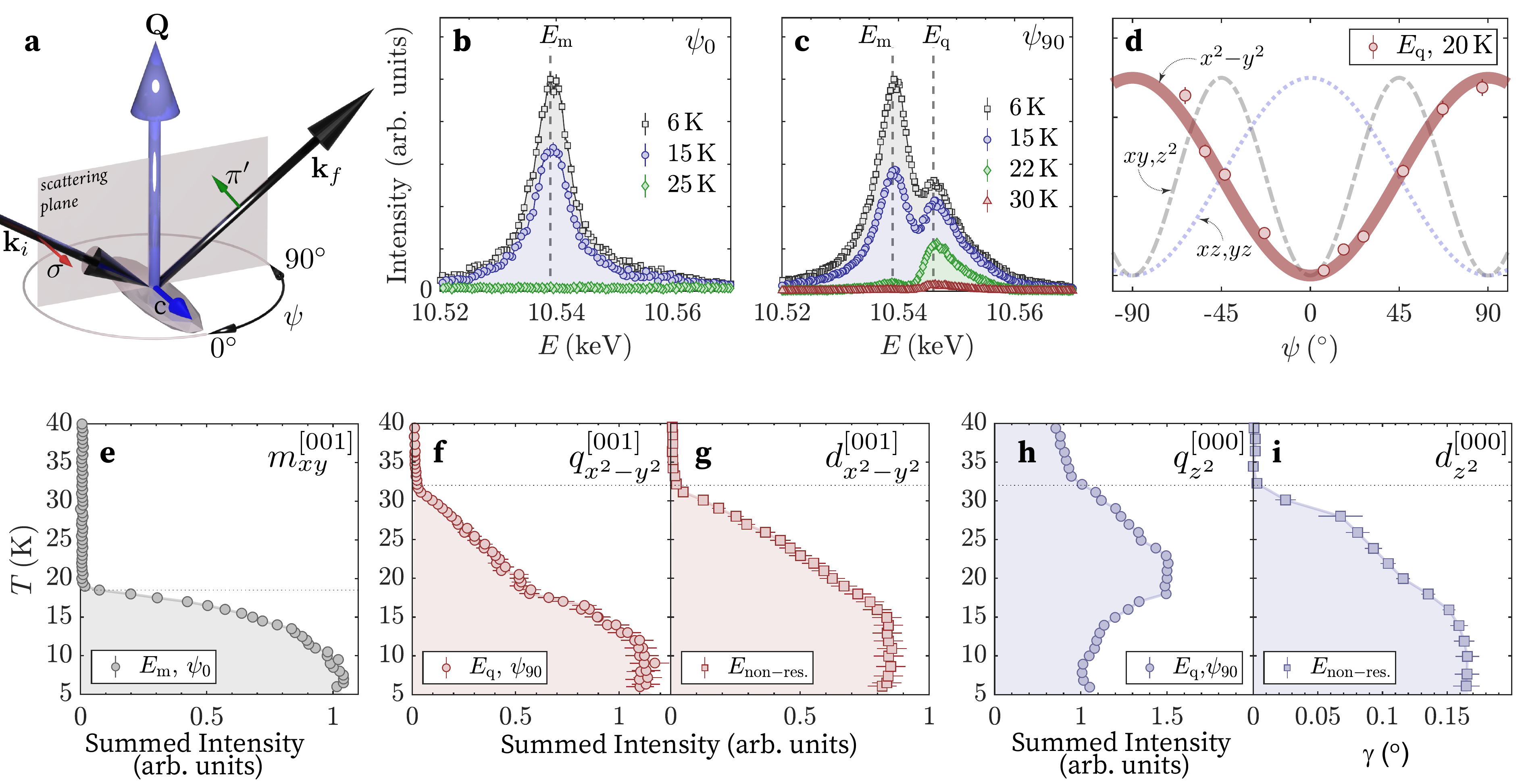}
\caption{\textbf{Direct detection of the charge quadrupolar, structural and magnetic order parameters in BMRO.} \textbf{a} REXS experimental setup in the $\sigma$$\to$$\pi^\prime$ scattering channel. The $\psi_0$ ($\psi_{90}$) orientation corresponds to the configuration where the crystal $c$ axis points perpendicular (within) the vertical scattering plane. \textbf{b,c} Energy dependence of the \textbf{Q}\,=\,($5,5,0$) Bragg reflection with the $\psi_0$ and $\psi_{90}$ crystal orientations, at various temperatures. \textbf{d} Azimuthal dependence of the ($5,5,0$) reflection at $E_\mathrm{q}$ and at 20\,K demonstrates the $\theta$=$x^2-y^2$ symmetry of the antiferroic quadrupolar $q_\theta^{[001]}$ order. The dashed (dotted) lines depict the calculated azimuthal dependence of the quadrupoles with $xy$,$z^2$ ($xz$,$yz$) symmetry. \textbf{e} Integrated intensity of the  ($5,5,0$) peak at the magnetic resonance $E_\mathrm{m}$ and $\psi_0$ azimuth indicates the onset of the antiferroic $m_{xy}^{[001]}$ order of Re magnetic dipoles below 18\,K. \textbf{f,g} Integrated intensity of the ($5,5,0$) and ($5,3,0$) peaks measured at $E_q$ and off-resonance, respectively, to detect the antiferroic $q_{x^2-y^2}^{[001]}$ and $d_{x^2-y^2}^{[001]}$ orders. \textbf{h,i} The ($10,0,0$) peak measured at $E_\mathrm{q}$ resonance and off-resonance to detect the ferroic $q_{z^2}^{[000]}$ and $d_{z^2}^{[000]}$ orders, respectively.}
\label{fig:FIG2}
\end{figure*}

As an example from the class of $5d^1$ DPs with their especially delicate interplay between electronic and lattice degrees of freedom, here we study \bmro{} (hereafter BMRO) with a combination of advanced theoretical and experimental techniques. BMRO features Re cations on a face-centered cubic (fcc) lattice in a formally 5$d^1$ electron configuration that is both strongly spin-orbit entangled and Jahn-Teller active due to its four-fold degenerate $J_\mathrm{eff}=3/2$ ground state~\cite{5d1_review_cooperation}.

First, using resonant and non-resonant elastic x-ray scattering, we present direct experimental evidence for the simultaneous long-range order of Re $5d^1$ electric quadrupoles ($q\kQ$) and structural distortions of the ReO$_6$ octahedra ($d\kQ$) below $T_\mathrm{q}$$\approx$$\SI{33}{K}$. Specifically, the $\theta$$=$$x^2$$-$$y^2$ mode exhibits an antiferroic arrangement with a propagation vector $\vk$$=$$[001]$, while the $\theta=z^2$ mode displays ferroic $\vk$$=$$[000]$ order [Figs.~\ref{fig:FIG1}\textbf{a},\textbf{b},\textbf{d},\textbf{e}].
Crucially, the detection of the concomitant onset of the $q^{[001]}_{x^2-y^2}$ and $q^{[000]}_{z^2}$ quadrupolar orders alongside the $d^{[001]}_{x^2-y^2}$ and $d^{[000]}_{z^2}$ structural distortions in BMRO represents a clear experimental verification of several theoretical studies~\cite{5d1_Mean_field_orbital_order, Chen2010, LoveseyBMRO, iwahara_vibronic_2023}.

Second, we show that calculations, where the role of the lattice distortions is neglected, predict two competing quadrupolar orders, namely the antiferroic $q_{x^2-y^2}^{[001]}$ order of the experimental ground state and a ferroic $q_{xy}^{[000]}$ order, which was not detected in the experiment. Both types of quadrupolar orders are stabilized by intersite electronic interactions alone. Coupling with lattice deformations then lowers the relative energy of the $q_{x^2-y^2}^{[001]}$ to the extent that it forms the ground state, in agreement with the experimental findings.

This departure from the conventional role of the lattice as a stabilizer of orders that are already favored electronically underscores the necessity for a more nuanced understanding of ordering transitions, where the cooperative contributions of electronic interactions and structural distortions shape the emergent properties of quantum materials. By elucidating the intricate interplay between these degrees of freedom, our work sheds light on the nature of the cooperative behavior of instabilities in complex materials and paves the way for new insights regarding emergent phenomena.

\section{Experimental Results} 
Prior experimental studies have reported two consecutive anomalies in the heat capacity of BMRO at temperatures $T_\mathrm{q}$\,=\,33\,K and $T_\mathrm{m}$\,=\,18\,K, demarcating three distinct phases~\cite{Hirai2019, Hirai2020, Pasztorova2022}: (i) above $T_\mathrm{q}$, in the paramagnetic-cubic phase, BMRO exhibits $Fm\bar{3}m$ symmetry with ReO$_6$ octahedra arranged on an fcc lattice; (ii) at $T_\mathrm{q}$, a cubic-to-tetragonal structural distortion occurs, leading to a transition into the paramagnetic-tetragonal phase with $P4_2/mnm$ space group symmetry; (iii) below $T_\mathrm{m}$, BMRO further develops long-ranged magnetic order and enters the magnetic-tetragonal phase. 

Although theoretical predictions have suggested the simultaneous onset of structural distortion and charge quadrupolar order in BMRO below $T_\mathrm{q}$~\cite{5d1_Mean_field_orbital_order, Chen2010, LoveseyBMRO, iwahara_vibronic_2023}, direct experimental evidence of quadrupolar order has not been demonstrated. Indeed, the preliminary x-ray scattering measurements of BMRO reported by Hirai~\textit{et al.}~\cite{Hirai2020} focused only on the magnetic and structural order. In this work, we extend the resonant elastic x-ray scattering (REXS) approach and detect an additional resonance at $E_\mathrm{q}$ ($\sim$10.541\,keV) that allows us to directly address the quadrupolar order developing at $T_\mathrm{q}$. 

Figs.~\ref{fig:FIG2}\textbf{b,c} displays the energy dependence of the \textbf{Q}=($5,5,0$) Bragg reflection across the rhenium $L_3$ absorption edge for two different azimuthal angles, $\psi_0 = 0^{\circ}$ and $\psi_{90} = 90^{\circ}$, as illustrated in Fig.~\ref{fig:FIG2}\textbf{a}. For the $\psi_0$ orientation at low temperatures ($T$=\,6\,K), we observe a single resonant peak centered at $E_\mathrm{m}$ ($\sim$10.535 keV), which is consistent with an earlier REXS study~\cite{Hirai2020}. On the other hand, the $\psi_{90}$ orientation reveals an additional resonance at $E_\mathrm{q}$, which has not been reported before. The energy scans performed at elevated temperatures in both configurations demonstrate distinct behaviors: the peak at $E_\mathrm{m}$ disappears upon heating above $T_\mathrm{m}$, whereas the peak at $E_\mathrm{q}$ persists well into the paramagnetic-tetragonal phase and vanishes at $T_\mathrm{q}$.

To gain a deeper understanding of the origins of the two resonances, we examine the temperature dependence of the ($5,5,0$) reflection with the incident photon energy fixed at $E_\mathrm{q}$ and $E_\mathrm{m}$, as shown in Figs.~\ref{fig:FIG2}\textbf{f} and \textbf{e}, respectively. The resonant peak at $E_\mathrm{m}$ displays a temperature dependence characteristic of an order parameter below $T_\mathrm{m}$, consistent with the anomaly observed in the magnetic susceptibility of BMRO~\cite{Hirai2019}. Thus, we attribute the resonance at $E_\mathrm{m}$ to the antiferromagnetic $\vk=[001]$ magnetic order of the Re sublattice, in line with the interpretation from the previously reported REXS study~\cite{Hirai2020}.

In contrast, the intensity of the ($5,5,0$) peak at the $E_\mathrm{q}$ resonance already exhibits a sharp increase below $T_\mathrm{q}$, while BMRO remains in the paramagnetic phase [Fig.~\ref{fig:FIG2}\textbf{f}]. As time-reversal symmetry is preserved, the observed REXS intensity must originate from the anisotropic tensor scattering from the antiferroic $\vk$=$[001]$ order of Re $5d^1$ charge quadrupoles~\cite{LoveseyBMRO}. This non-zero intensity of the ($5,5,0$) Bragg peak at $E_\mathrm{q}$ within the temperature range between $T_\mathrm{m}$ and $T_\mathrm{q}$ provides direct evidence for the long-range order of charge quadrupoles in the $5d^1$ double perovskite BMRO.

We perform a symmetry analysis of the azimuthal dependence of the ($5,5,0$) Bragg peak at $E_\mathrm{q}$ in the paramagnetic tetragonal phase ($T$\,=\,25\,K), as shown in Fig.~\ref{fig:FIG2}\textbf{d}. The data provides a consistent fit with antiferroic ordering of charge quadrupoles possessing $x^2-y^2$ symmetry, namely $q_{x^2-y^2}^{[001]}$. Crucially, our analysis also excludes $q_{\theta}^{[001]}$ orders with $\theta$$=$$z^2$, $xy$, $xz$ or $yz$ symmetry, as shown by the broken lines in Fig.~\ref{fig:FIG2}\textbf{d}.

Furthermore, the emergence of the $q^{[001]}_{x^2-y^2}$ quadrupolar order below $T_\mathrm{q}$ occurs concomitantly with the onset of the $d^{[001 ]}_{x^2-y^2}$ structural distortion, as shown by the temperature dependence of the ($5,3,0$) reflection measured at $E_\mathrm{non-res.}$ (10.500 keV) [Fig.~\ref{fig:FIG2}\textbf{g}]. This structural peak, which is otherwise forbidden by the $Fm\bar{3}m$ cubic structure, arises from an antiferroic $\vk$=[001] ordering of $x^2-y^2$ distortions of the ReO$_6$ octahedron as illustrated in Fig.~\ref{fig:FIG1}\textbf{d}.

In addition to the antiferroic $q^{[001]}_{x^2-y^2}$ and $d^{[001]}_{x^2-y^2}$ orders, we also found evidence for the spontaneous ferroic $\vk$=[000] long-range ordering of $z^2$-type lattice distortion and charge quadrupoles below $T_\mathrm{q}$ [Fig.~\ref{fig:FIG1}\textbf{e}]. The former, $d^{[000]}_{z^2}$, manifests as a cubic-to-tetragonal structural distortion [Fig.~\ref{fig:FIG1}\textbf{e}] as shown by the splitting of the ($10,0,0$) structural peak in Fig.~\ref{fig:FIG2}\textbf{i}. On the other hand, discerning the experimental signature of the latter is more challenging as the scattered intensity from the $q^{[000]}_{z^2}$ order occurs at the same reciprocal space location as the structural peaks. To maximize the ferroquadrupolar signal, we use incident x-ray energy at $E_\mathrm{q}$ to benefit from REXS, in conjunction with scattering in the $\sigma$$\to$$\pi^\prime$ channel [Fig.~\ref{fig:FIG2}\textbf{a}] to suppress the signal from the structural peak. We are then able to observe a sharp increase in the scattering intensity just below $T_\mathrm{q}$ [Fig.~\ref{fig:FIG2}\textbf{h}], consistent with the onset of $q^{[000]}_{z^2}$ before the signal eventually decreases due to the splitting of the underlying structural peak.

\begin{figure*}[t!]
\centering
\includegraphics[width=1.99\columnwidth]{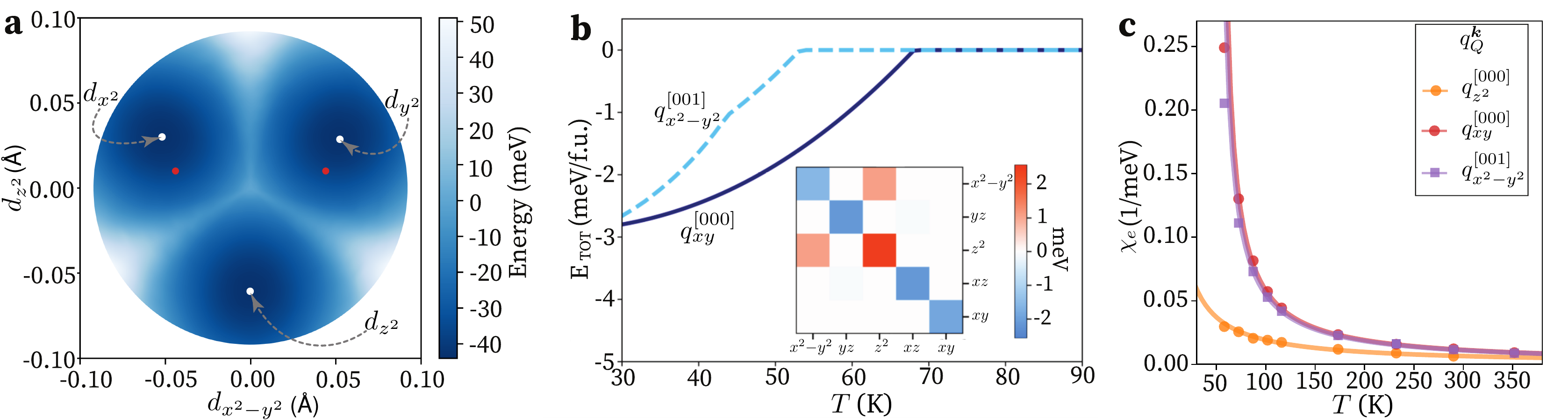}
\caption{\textbf{Calculations of BMRO.} \textbf{a} The potential energy surface of a single ReO$_6$ octahedron in BMRO as a function of $d_{x^2-y^2}$ and $d_{z^2}$ distortions, compared to the experimental octahedral distortions (denoted with the red dots) \cite{Hirai2020}. \textbf{b} Total energy $E_\mathrm{TOT}$ over the temperature $T$ calculated within the mean-field approximation from the \textit{ab-initio} IEI Hamiltonian (Eq.~(\ref{eq:H_IEI})). The solid and dashed curves are the total energies of the $t_{2g}$ ferro-$q_{xy}^{[000]}$ and $e_g$ antiferro-$q_{x^2-y^2}^{[001]}$ orders, respectively. Inset: the quadrupole-quadrupole block of the IEI matrix $V$ for the nearest-neighbor $xy$ Re-Re bond, blue (red) colors correspond to ferro (antiferro) couplings. \textbf{c} The electronic susceptibility $\chi_\mathrm{e}$ over the temperature $T$ associated with the ferro $xy$ and antiferroic $\vk=[001]$ $x^2-y^2$ quadrupolar orders display Curie-Weiss behaviour with a $T_\mathrm{q}\sim$\,50\,K. On the other hand, the $z^2$ quadrupoles do not diverge at a temperature above zero and thus do not ferroically order spontaneously.}
\label{fig:FIG3}
\end{figure*}

\section{Computational Results}
We now seek to understand the driving force behind the antiferroic $x^2-y^2$ and ferroic $z^2$ ordering [Figs.~\ref{fig:FIG1}\textbf{d}, \textbf{e}] and its simultaneous onset in both the lattice and electronic degrees of freedom in the paramagnetic phase at $T_\mathrm{q}$. To that end, we separate computationally the effect of the lattice instabilities of the ReO$_6$ octahedron from the intersite electronic interaction and investigate the resulting order parameters ($q_\theta^\vk$, $d_\theta^\vk$). 

Through this comparison, we will demonstrate that (i) the local Jahn-Teller effect or (ii) the electronic effects separately cannot explain the experimentally observed orders. Only when (iii) both effects are accounted for can the ground state in BMRO be established. Hence, in the following sections, we examine these three scenarios (i)--(iii). In particular, we demonstrate that while the quadrupolar transition at $T_\mathrm{q}$ is driven by the electronic interactions, the coupling to the lattice distortion $d^{[001]}_{x^2-y^2}$ is crucial to obtain the experimentally observed antiferroic $q_{x^2-y^2}^{[001]}$ order. Moreover, we show that the coupled ferroic $d^{[000]}_{z^2}$ and $q^{[000]}_{z^2}$ order is secondary.

\subsection{(i) Local Jahn-Teller effect}

We first investigate the role of the local Jahn-Teller effect by determining what kind of structural deformations ($d_\theta$) has the propensity to lower the energy of BMRO. In order to go beyond isolated octahedral models~\cite{bersuker_vibronic_1989, streltsov_interplay_2022}, these distortions were applied to a ReO$_6$ octahedron embedded within the BMRO lattice. In this work, the potential energy landscape with respect to these distortion modes were determined by means of quantum-chemical (multi-reference configuration interaction) calculations. 


Our calculations reveal a threefold degenerate global minima at \SI{-44}{meV}, which is established by $e_g$-type distortions alone (namely $d_{x^2-y^2}$ and $d_{z^2}$ [Figs.~\ref{fig:FIG1}\textbf{a},\textbf{b}]). As shown in the energy landscape in Fig.~\ref{fig:FIG3}\textbf{a}, these minima reside at $d_{z^2} = \SI{-.060}{\angstrom}$, 
along with its symmetry equivalents $d_{x^2}$ and $d_{y^2}$.
On the other hand, $t_{2g}$-type distortions alone (namely $d_{xy}$ [Fig.~\ref{fig:FIG1}\textbf{c}], $d_{yz}$ and $d_{xz}$) can only lower the energy by \SI{12}{meV}, at most, relative to the undistorted ReO$_6$ state and hence do not realize the ground state of the system.

Thus, the consideration of purely local effects alone can predict the same type of lattice distortions that was detected in the experiment, namely those of $e_g$ symmetry. In particular, an alternating arrangement of $d_{x^2}$$<$$0$ and $d_{y^2}$$<$$0$ distortions, equivalent to $\pm (\sqrt3/2) d_{x^2-y^2} + (1/2) d_{z^2}$, is qualitatively consistent with the experimental distortion amplitudes~\cite{Hirai2020}, as indicated by the red dots in Fig.~\ref{fig:FIG3}\textbf{a}. Nonetheless, a purely local model comprising a single ReO$_6$ octahedron is insufficient in predicting how the local distortions are arranged across the lattice, to form long-range order.

\subsection{(ii) Purely electronic interactions}
Next, to study the role of the electronic effects and determine what kind of long-range electronic order ($q_\theta^\vk$) might arise, we calculate the intersite exchange interactions (IEI). To isolate the effect of the IEI from the local Jahn-Teller coupling, we consider the undistorted, cubic lattice of BMRO. The IEI are predicted to be the main ordering mechanism of many $5d^1$ and $5d^2$ DPs~\cite{5d1_Mean_field_orbital_order, Chen2010, 5d2_chen_gang} and come from the virtual hopping of the $5d$ electrons through ligands. In this work, they are calculated using the force theorem implemented within the Hubbard-I (FT-HI) framework~\cite{Pourovskii2016}. 

The IEI can be written as couplings between local multipole operators $O_{K}^\theta(i)$ \cite{Santini2009, Pourovskii2016} of order $K$ and momentum projection $\theta$ at the Re site $i$. The multipole operators provide a complete representation of the electronic degrees of freedom in the Re-$J_\mathrm{eff} = 3/2$ manifold, where $K=1$, 2, and 3 label the dipole, quadrupole and octupole moments, respectively.
Our computed IEI for a given Re-Re bond $\langle ij\rangle$ can then be expressed as the matrix $V_{KK'}^{\theta \theta'}(ij)$, and  the full IEI Hamiltonian reads
\begin{equation}
    H_\mathrm{IEI}=\sum_{\langle ij\rangle} \sum_{K\theta K'\theta'} V_{KK'}^{\theta \theta'}(ij) O_{K}^\theta (i) O_{K'}^{\theta'}(j) ,
    \label{eq:H_IEI}
\end{equation}
where the first summation is over all nearest-neighbor Re-Re bonds, see Methods. The operators labeled with $\theta$ ($\theta'$) act on $i$ ($j$) site, respectively.

The inset of Fig.~\ref{fig:FIG3}\textbf{b} shows the calculated quadrupole-quadrupole IEI, which dominate over dipole and octupole terms, for a pair of Re atoms in the $xy$ plane. While the interaction between the $t_{2g}$
($xy$, $xz$, $yz$) quadrupoles is negative, indicating the tendency towards a ferroic order, the strongest $e_g$ ($x^2-y^2$, $z^2$) quadrupole coupling is positive and thus antiferroic. Thus, the ferroic $t_{2g}$ order is expected to be favored over antiferroic orders, due to the geometrical frustration of the fcc lattice. 

Based on these IEI, we identify the stable electronic orderings $q^\vk_\theta$ by numerically solving Eq.~\ref{eq:H_IEI} within the mean-field approximation. The resulting total energy as a function of the temperature is shown in Fig.~\ref{fig:FIG3}\textbf{b}.
A ferro $q^{[000]}_{xy}$ quadrupolar ordered state emerges with the highest ordering temperature $T_\mathrm{q} = \SI{68}{K}$.
The competing phase is identified by solving  (\ref{eq:H_IEI}) with the IEI between $t_{2g}$ quadrupoles suppressed. In this case, the experimentally observed antiferroic $q_{x^2-y^2}^{[001]}$ order emerges at a slightly lower $T_\mathrm{q}$$=$$\SI{53}{K}$, indicating that the antiferroic $e_g$ order is less stable than the ferroic $t_{2g}$ quadrupolar order in this purely electronic model. Our mean-field calculations with the FT-HI-derived IEI overestimate the transition temperature $T_\mathrm{q}$, which is expected due to the mean-field approximation and is in line with previous studies of DPs~\cite{Pourovskii_d2,d3}. 

We complement these results on the electronic order with density-functional theory plus dynamical mean-field theory (DFT+DMFT) calculations that go beyond the quasi-atomic Hubbard-I approximation, again for the cubic, paramagnetic phase without lattice distortions.
We compute the electronic susceptibilities $\chi_\mathrm{e}(T)$ of cubic BMRO under small (anti-)ferroic quadrupolar fields~\cite{schaufelberger_exploring_2023} as a function of the temperature $T$, which are then fitted with a Curie-Weiss law, $\chi_\mathrm{e}(T) \propto (T-T_\mathrm{q})^{-1}$.

As shown in Fig.~\ref{fig:FIG3}\textbf{c}, $\chi_\mathrm{e}$ diverges for both the $q_{xy}^{[000]}$ and $q_{x^2-y^2}^{[001]}$ orders. Consistent with the FT-HI results, this indicates the emergence of electronic quadrupolar long-range order at around \SI{50}{K}, with a slightly higher $T_\mathrm{q}$ for $q_{xy}^{[000]}$. The $q_{z^2}^{[000]}$ order, which also appeared in our REXS measurements, does not exhibit a divergence for $T>0$, which indicates its inability to spontaneously order.

Together, the FT-HI and DFT+DMFT calculations with the lattice frozen to its high-symmetry structure draw a consistent picture, demonstrating that electronic interactions alone can stabilize long-range quadrupolar order.
However, both calculations (where the role of the Jahn-Teller effect is neglected) indicate a slightly stronger tendency towards spontaneous ferro $q_{xy}^{[000]}$ quadrupolar order, contrary to the experimentally observed $q_{x^2-y^2}^{[001]}$ order. 

Moreover, these predictions 
are inconsistent with the case 
when only considering local Jahn-Teller effects discussed in the previous section, where $d_{x^2-y^2}$ and $d_{z^2}$ local distortions are preferred over $d_{xy}$. This contradiction clearly indicates that models that only consider the effects of IEI or local Jahn-Teller coupling separately are insufficient to comprehensively describe the emerging long-range order in BMRO. 

\subsection{(iii) Combined Jahn-Teller and electronic model}

Hence, we now combine the Jahn-Teller and electronic effects discussed before within a single computational framework. This is achieved by coupling the $q_{x^2-y^2}^{[001]}$, $q_{xy}^{[000]}$, and $q_{z^2}^{[000]}$ quadrupolar orderings to the $d_{x^2-y^2}^{[001]}$, $d_{xy}^{[000]}$, and $d_{z^2}^{[000]}$ distortion modes, respectively. For the two $\vk$=[000] modes, both homogeneous strain and internal oxygen displacements distort the ReO$_6$ octahedra in the same way, which we account for by defining the corresponding distortion amplitudes as two-component vectors $\bm d_\theta^\vk$.

The leading order contribution of the structural deformation modes to the Landau-type free energy can be described as:
\begin{align}
    F_\mathrm{latt}(q_\theta^\vk, \bm d_\theta^\vk) = - \frac12 \bm g_\theta^\vk \cdot \bm d_\theta^\vk q_\theta^\vk + \frac12 \bm d_\theta^\vk \cdot( \bm K_\theta^\vk \bm d_\theta^\vk) ,
    \label{eq:latt_model}
\end{align}
where $\bm g_\theta^\vk$ defines the bilinear coupling strength between the distortion $\bm d_\theta^\vk$ and the quadrupole $q_\theta^\vk$ of the same symmetry and the matrix $\bm K_\theta^\vk$ parametrizes the energy cost for elastic lattice deformations \cite{peil_mechanism_2019, georgescu_disentangling_2019, georgescu_quantifying_2022}.
A consequence of the bilinear electron-lattice coupling is that the quadrupole moments directly couple to distortions with
\begin{align}
    \bm d_\theta^\vk = \frac12 (\bm K_\theta^\vk)^{-1} \cdot \bm g_\theta^\vk q_\theta^\vk ,
    \label{eq:equi_ampl}
\end{align}
which follows from minimizing the free energy with respect to $\bm d_\theta^\vk$. This direct connection between the electronic and lattice degrees of freedom in Eq.~\ref{eq:equi_ampl} explains the simultaneous and proportional onset of their long-range ordering at $T_\mathrm{q}$, as observed in our experiments [Fig.~\ref{fig:FIG2}\textbf{f}--\textbf{i}].

\begin{table}[bt]
\caption{The lattice stabilization energy can be obtained from the electron-lattice coupling $\bm g$ and the elastic deformation $\bm K$ through a simple Landau free-energy expansion.}
\label{table1}
\begin{ruledtabular}
\begin{tabular}{cccccc}
$\bm k$ & $\theta$ & $\bm g$ & $\bm K$ & $F_\mathrm{latt}^\mathrm{sat}$ \\
&  & (eV/\AA) & (eV/\AA$^2$) & (meV/f.u.) \\\hline\\[0.1pt]
[001] & $x^2-y^2$ &  $2.52$ & $19.9$ & -10.0 \\[8pt] 
[000]  &$xy$ &  $\begin{pmatrix} 0.76 \\ 0.92 \end{pmatrix}$ & 
$\begin{pmatrix} 9.5 & 3.8 \\ 3.8 & 21.8 \end{pmatrix}$ & -2.5 \\[10pt]
[000] & $z^2$ &  $\begin{pmatrix} 1.93 \\ 2.48 \end{pmatrix}$ & 
$\begin{pmatrix} 18.4 & 13.1 \\ 13.1 & 21.1 \end{pmatrix}$ & -9.6 \\
\end{tabular}
\end{ruledtabular}
\end{table}

At low temperatures, where the quadrupolar order is fully saturated ($|q_\theta^\vk| = 1/2$), the free energy in Eq.~\ref{eq:latt_model} can be simplified to give
$F_\mathrm{latt}^\mathrm{sat} = -\bm g_\theta^\vk \cdot ((\bm K_\theta^\vk)^{-1} \bm g_\theta^\vk) / 32$. Now, to ascertain which of the three types of order acquires the most stability, we compute the material constants $\bm g_\theta^\vk$ and $\bm K_\theta^\vk$ using DFT, so as to obtain the associated $F_\mathrm{latt}^\mathrm{sat}$ term (Table~\ref{table1}). 
%

As shown in the last column of Table~\ref{table1}, the $F_\mathrm{latt}^\mathrm{sat}$ term stabilizes the experimentally-observed $q_{x^2-y^2}^{[001]}$ order much more strongly compared to the $q_{xy}^{[000]}$ one, by a factor of 4. Crucially, when compared to the energy scales involved in Fig.~\ref{fig:FIG3}\textbf{b}, the difference in $F_\mathrm{latt}^\mathrm{sat}$ between the two orderings is strong enough to change their relative stability, to the extent of establishing the $q_{x^2-y^2}^{[001]}$ order as the ground state, in accordance with the experimental findings. We note that the $F_\mathrm{latt}^\mathrm{sat}$ term also endows the $q_{z^2}^{[000]}$ order with additional stability, but is eventually not strong enough to compete with the $q_{x^2-y^2}^{[001]}$ order.

Our calculations demonstrate that, while charge quadrupolar order is energy lowering in the absence of lattice distortions (as shown in section ii), the electron-lattice coupling is crucial to favouring the experimentally-observed order as the ground state, relative to other types of quadrupolar order. 

\section{Summary}
In summary, we provide convincing proof for long-ranged antiferroic order of Re $5d^1$ charge quadrupoles below $T_\mathrm{q}$ with a propagation vector of $\vk=[001]$. Our analysis demonstrates that only the Re $5d^1$ quadrupoles that transform with $\theta = {x^2-y^2}$ symmetry can account for the azimuthal dependence of the corresponding resonant elastic x-ray scattering intensity. Furthermore, we demonstrate that this $q_{x^2-y^2}^{[001]}$ order occurs concomitantly with the $q_{z^2}^{[000]}$ quadrupolar order and the $d_{x^2-y^2}^{[001]}$ and $d_{z^2}^{[000]}$ structural distortions. 

Our comprehensive computational modeling of BMRO demonstrates that the long-range ordering of the quadrupoles ($q_\theta^{\vk}$) and lattice distortion ($d_{\theta}^{\vk}$)  of  $\theta$$=$$x^2-y^2$ symmetry is the primary order parameter, whereas that of $\theta$$=$$z^2$ symmetry is secondary. Moreover, we demonstrate that while the $q_{x^2-y^2}^{[001]}$ order occurs spontaneously due to electronic interactions, it is the coupling to the lattice distortion of $d_{x^2-y^2}^{[001]}$ type that establishes it as the ground state over the competing $q_{xy}^{[000]}$ ferroic order.

\section*{Methods}
\textbf{Crystal growth and bulk characterization.} The BMRO single crystals were synthesized via the flux growth method outlined in Refs.~\cite{Hirai2019, Hirai2020}.  In this process, a mixture of BaO, MgO, and ReO$_3$ powders was combined with a flux comprising 36 wt\% BaCl$_2$ and 64 wt\% MgCl$_2$ within an argon-filled glove box. The resulting mixture was then sealed in a platinum tube and subjected to a gradual temperature increase up to 1300$^\circ$C. Subsequently, the tube was slowly cooled at a rate of 5$^\circ$C/h, reaching a final temperature of 900$^\circ$C. Once the system reached room temperature through natural cooling, the remaining flux was thoroughly eliminated by rinsing with distilled water. This process yielded octahedral-shaped BMRO single crystals with typical dimensions of approximately 1$\times$1$\times$1 mm$^3$. The structure and quality of the obtained single crystals were verified using a laboratory 6-circle x-ray diffractometer (Rigaku).
	
\textbf{Resonant elastic X-ray scattering.} We performed low-temperature resonant elastic x-ray scattering (REXS) at the P09 beamline at PETRA III (DESY)~\cite{Strempfer} using a closed cycle cryostat. The crystal was aligned with the $[110]$ axis in the vertical scattering plane, to study the ($h$,$h$,$0$) reflections in the specular diffraction condition. The angle $\psi$ is defined as the rotation angle of the crystal about the scattering vector, $\textbf{Q}$. As such, the angle $\psi_0=0^\circ$ ($\psi_{90}=90^\circ$) refers to the sample configuration in which the crystal $[001]$ direction is perpendicular to (within) the vertical scattering plane. A pyrolytic graphite (002) analyzer crystal was used to discriminate between the $\sigma^\prime$ and $\pi^\prime$ scattered x-ray linear polarisations, which lie perpendicular and within the vertical scattering plane, respectively.

\textbf{Quantum chemistry calculations.}
Embedding multi-reference quantum chemistry calculations were performed using the Molpro package~\cite{Molpro}. The crystal structure for $Fm\bar{3}m$ configuration of BMRO was taken from Ref.~\cite{Hirai2020}.

The quantum cluster under consideration consisted of 21 atoms: the ReO$_6$ octahedron with the 6 nearest Mg atoms and eight nearest Ba atoms. For the Re atom, core potentials and a basis set consisting of triple-zeta functions plus two polarization $f$ functions were used ~\cite{ReBasis}. All-electron triple-zeta functions were used for the surrounding O atoms ~\cite{OBasis}. Both Mg and Ba atoms in the cluster were described by an effective core potential and supplemented with a single $s$ function~\cite{MgBasis1,MgBasis2}. The crystal lattice in which the cluster was embedded was constructed at the level of a Madelung ionic potential, with formal charges placed at the ionic sites. Beyond the first 500 ions, these formal charges were changed such that the unmodified direct sum potential matched the unmodified Ewald summation, as described in Ref.~\cite{Ewald}.

For the complete active space self-consistent field (CASSCF) calculations, an active space of 5 orbitals ($3 t_{2g} + 2 e_g$) was used. The optimization was carried out for an average of 5 states (ground state + 4 excited states) of the scalar relativistic Hamiltonian with each state having a weight of one. Multi-reference configuration interaction (MRCI) treatment was performed with single and double substitutions with respect to the CASSCF reference, as described in
Refs.~\cite{MRCI1,MRCI2}. Spin-orbit coupling was included as described in Ref.~\cite{SOC}.

The MRCI calculations were performed on the following grid: Isotropic distortions were varied from 0.5\% to -1.5\% with 0.5\% steps. In this work, we show the results for an isotropic compression of \SI{1}{\%}, for which the global energy minimum was found. For each value of isotropic distortion, a \ang{60} sector of the $d_{z^2}$-$d_{x^2-y^2}$ plane was simulated, with angle being varied with steps of \ang{15} and radius being varied from 0 to 1.4\%, with step of 0.1\% (with \% referencing the deviation of $x$-axis atoms from unperturbed \SI{1.926}{\angstrom} Re-O interatomic distance). The rest of the $d_{z^2}$-$d_{x^2-y^2}$ plane was filled exploiting the symmetry of the $d_{x^2-y^2}$ amplitude and the 3-fold symmetry of the $d_{z^2}$-$d_{x^2-y^2}$ plane.

In the investigation of the $t_{2g}$ modes, the isotropic and $d_{yz}+d_{xz}+d_{xy}$ as well as the $d_{z^2}$ and $d_{x^2-y^2}$ deformation modes were taken as orthogonal minimization axes to capture all (local) minima predicted by octahedral models \cite{bersuker_vibronic_1989, streltsov_interplay_2022}. Starting with the undistorted $Fm\bar{3}m$ structure, the search for minima was conducted by iterative optimization along these axes. Several minimization paths were considered that all converged back to the $d_{z^2}<0$ minimum mentioned in the results section.

\textbf{FT-HI calculations of exchange interactions.}
To calculate the IEI by the FT-HI method, we started with self-consistent DFT+DMFT calculations using the HI approximation to solve the Re 5$d$ impurity problem. Our full-potential DFT+DMFT approach is based on the Wien2k DFT code~\cite{Wien2k} and ``TRIQS" library implementation of DMFT~\cite{triqs_main,triqs_dft_tools}. In these calculations, we included spin-orbit coupling and used the experimental BMRO lattice structure, local-density approximation (LDA) exchange-correlation, 400 $\vk$-points in the full Brillouin zone, as well as the Wien2k basis cutoff $R_{\mathrm{mt}}K_{\mathrm{max}}$=7. The projective Wannier orbitals~\cite{Amadon2008, Aichhorn2009} for Re $t_{2g}$ orbitals are constructed from the corresponding manifold of $t_{2g}$-like bands. The rotationally invariant Kanamori interaction Hamiltonian was constructed using the parameters $U^\mathrm{K}$=3.77\,eV and $J_\mathrm{H}^\mathrm{K}$=$0.39$\,eV; the corresponding interaction parameters for the full-5$d$ shell are $U$=$F^0$=3.2\,eV and $J_\mathrm{H}$=0.5\,eV in agreement with previous works~\cite{5d1_DMFT_orbital_order, Pourovskii_d2}. After DFT+HI calculations are converged, the FT-HI method \cite{Pourovskii2016} was applied as post-processing on top of the converged DFT+HI electronic structure. We find, in agreement with previous calculations of IEI in 5$d$ double perovskites~\cite{5d1_DMFT_orbital_order, Pourovskii_d2, d3}, that due to a rapid decay of superexchange with the distance, only the nearest-neighbor IEI terms are important. We used the ``McPhase" \cite{Rotter2004} package with an in-house module to solve the IEI Hamiltonian in mean field.

\textbf{DFT+DMFT calculations.}
To calculate the quadrupolar susceptibility, we performed paramagnetic DFT+DMFT calculations, where the local impurity problem is solved using a numerically exact Quantum Monte Carlo method, which incorporates the hybridization with the effective bath.
For the DFT calculations, we used VASP (version 6.4.1) \cite{kresse_ab_1993, kresse_efficient_1996} with the PBE PAW pseudopotentials \cite{perdew_generalized_1996} including the Ba-$5s$, Ba-$5p$, Mg-$2p$, and Re-$5p$ semicore states as valence electrons. The calculations were performed in a 20-atom unit cell of the experimental, cubic $Fm\bar3m$ structure \cite{Hirai2020} with a $7\times7\times5$ reciprocal grid and a \SI{600}{eV} plane-wave cutoff down to an energy convergence of \SI{e-8}{eV}. We included spin-orbit coupling and constrain the magnetic moments to zero to enforce time-reversal symmetry.
We generated Wannier orbitals from projections without localization with Wannier90 (version 3.1.0) \cite{pizzi_wannier90_2020} of the $t_{2g}$-like bands around the Fermi energy. The quadrupole operators $O_2^\theta$ are upfolded from the $J_\mathrm{eff} = 3/2$ basis to this Wannier basis.
We then ran one-shot DMFT calculations with solid\_dmft (version 3.1.0) \cite{merkel_solid_dmft_2022} (modified to implement the real spin-orbit coupled basis $\ket{xz}$, $\ket{-\mathrm i yz}$, $\ket{\mathrm i xy}$ and the external field) using the CT-HYB Quantum-Monte-Carlo impurity solver \cite{seth_triqscthyb_2016}, all from the TRIQS library \cite{triqs_main, Aichhorn2009}. We employed the same interaction parameters $U^\mathrm{K}$ and $J^\mathrm{K}_\mathrm{H}$ as in FT-HI and time-reversal symmetrize the impurity Green's function from the solver to ensure a paramagnetic solution.
The two Re sites in the unit cell were mapped onto each other by either the identity (for $q^{[000]}_{xy}$ and $q^{[000]}_{z^2}$) or a \ang{90} rotation about the $z$ axis (for $q^{[001]}_{x^2-y^2}$) so that the two Re sites in the unit cell are symmetry equivalent.
We applied an external field coupling to the different quadrupolar orders, which is defined as a local potential $\delta V = -s\kQ O_2^\theta$ \cite{schaufelberger_exploring_2023}, with a small $s\kQ = \SI{1}{meV}$ to ensure a linear, convergeable response. We then fitted the inverse susceptibility $(\chi\kQ)^{-1} = s\kQ / q_\theta^\vk$ with a linear function according to the Curie-Weiss law. Below an electronic temperature of \SI{50}{K}, the average sign from the impurity solver decays rapidly to zero and eventually makes convergence impossible.

\textbf{Lattice-related constants}.
To extract the material constants $\bm g_\theta^\vk$ and $\bm K_\theta^\vk$, we applied small distortions $d_{x^2-y^2}^{[001]}$, $d_{xy}^{[000]}$, and $d_{z^2}^{[000]}$ with the same symmetry as the previously discussed quadrupolar orders, or equivalently, $X_2^+$, $\Gamma_5^+$, and $\Gamma_3^+$ as defined in ISODISTORT \cite{campbell_isodisplace_2006, stokes_isodistort_2022}. 
We performed DFT calculations in the same setup as for the DFT+DMFT calculations, just without spin-orbit coupling, and get the total energy $E_\mathrm{DFT}$ as well as the matrices in orbital space of the local potential $\bm\epsilon$ and the local occupations $\bm\rho$ as a function of $\bm d\kQ$.
To calculate $\bm K\kQ$, we followed Refs.~\cite{peil_mechanism_2019, georgescu_disentangling_2019}, extending the formalism to the multidimensional distortions and orbital-dependent quantities. We fitted the DFT energy for different mode amplitudes with a quadratic polynomial $E_\mathrm{DFT} (\bm d_\theta^\vk) = \bm d_\theta^\vk \cdot( \bm C_\theta^\vk \bm d_\theta^\vk)/2$, where $\bm C\kQ$ is a matrix. This energy is the sum of the mode stiffness energy $\bm d_\theta^\vk \cdot( \bm K_\theta^\vk \bm d_\theta^\vk)/2$ from Eq.~\ref{eq:latt_model} and the DFT local occupation energy $\Tr(\bm\rho_\mathrm{DFT} \bm\epsilon)$ \cite{georgescu_disentangling_2019}.
If we take the derivative with respect to $\bm d_\theta^\vk$ and use the Hellmann-Feynman theorem, we obtain
\begin{align}
    \bm C_\theta^\vk \bm d_\theta^\vk = \frac{\partial E_\mathrm{DFT}}{\partial \bm d_\theta^\vk} = \bm K_\theta^\vk \bm d_\theta^\vk + \sum_{mn} \rho_{mn} \frac{\partial \epsilon_{nm}}{\partial \bm d_\theta^\vk} ,
\end{align}
and can therefore calculate $\bm K\kQ$.
$\bm g_\theta^\vk$ can directly be computed from the local potential. If only the $J_\mathrm{eff} = 3/2$ quadruplet is occupied, as is the case in the DFT+DMFT calculations with spin-orbit coupling, we can rewrite $\Tr(\bm \rho \bm \epsilon) = \Tr(\bm \rho)\Tr(\bm \epsilon) - \sum_\theta \Tr(-\bm \epsilon O_2^\theta)\Tr(\bm \rho O_2^\theta)$ for any time-reversal-symmetric occupation $\rho$. The first term is a mode-independent energy and the second term is the electron-lattice coupling that we can expand in leading order as $\Tr(-\bm \epsilon O_2^\theta) = \bm g_\theta^\vk \cdot \bm d_\theta^\vk / 2$ so that the coupling strength in Eq.~\ref{eq:latt_model} is a linear fit to the multipole decomposition of $\bm\epsilon$ over $\bm d_\theta^\vk$.

\section*{Data availability}
The data presented in this study is available upon reasonable request. Data about the DFT+DMFT calculations and the lattice-related constants will be released before publication.

\bibliographystyle{unsrt}
\bibliography{ref34}

\section*{Acknowledgments}
We thank Dan Porter, Chen Gang, Leon Balents, George Jackeli, Pablo J. Bereciartua Perez, Claude Monney, Yikai Yang, Thorsten Schmitt for fruitful discussions, Sophie Beck for the implementation of the real formulation of spin-orbit coupling in DMFT in solid\_dmft, and Christian Plueckthun for technical help.
We acknowledge DESY (Hamburg, Germany), a member of the Helmholtz Association HGF, for the provision of experimental facilities. Beamtime was allocated for proposal I-20211583 EC. Quantum chemistry calculations were performed at the facilities of Scientific IT and Application Support Center of EPFL. This work is supported by the ERC Synergy grant HERO under the Grant \# 810451 (J.R.S., H.M.R., I.Z., N.A.S., A.U., J.P.), Swiss National Supercomputing Centre under project grant ID s1128 (M.E.M., C.E.), ETH Z\"urich (M.E.M., C.E., A.U., N.A.S.) and the Singapore National Science Scholarship from the Agency for Science Technology and Research (J.R.S.). L.V.P. acknowledges the support of the computer team at CPHT.

\section*{Author contributions}
J.R.S. and H.M.R. designed the study. D.H. synthesized the single crystals. J.R.S., J.P., D.H., I.Z., and D.T. characterized the single crystals. J.R.S., J.P., S.F., I.Z., and D.T. performed the REXS experiments.
M.E.M., L.V.P., O.M., and D.F.M. carried out the theoretical calculations.
J.R.S., M.E.M., C.E., L.V.P., and O.M. wrote the manuscript with contributions from all authors.

\section*{Competing interests}
The authors declare no competing interests.

\section*{Additional information}
Correspondence and requests for materials should be addressed to Jian-Rui Soh.

\end{document}